\documentclass[useAMS,usenatbib]{mn2e}
\usepackage{graphicx}
\title[Australia Telescope observations of 9P/Tempel 1]
{Radio observations of comet 9P/Tempel 1 
with the Australia Telescope facilities during
the Deep Impact encounter.}
\author[P. A. Jones et al.]{P. A. Jones$^{1}$\thanks{E-mail:Paul.Jones@csiro.au}, 
J. M. Sarkissian$^{1}$, M. G. Burton$^{2}$,  M. A. Voronkov$^{1,3}$
and M. D. Filipovi{\'c}$^{4}$\\
$^{1}$Australia Telescope National Facility, PO Box 76, Epping NSW, 1710,
Australia\\
$^{2}$School of Physics, University of New South Wales, Sydney 2052, 
Australia \\
$^{3}$Astro Space Centre, Profsouznaya st. 84/32, 117997 Moscow, Russia \\
$^{4}$University of Western Sydney, Locked Bag 1797, Penrith South, DC, 
NSW 1797, Australia }

\begin{document}

\date{Accepted . Received ; in original form 2005 Nov 23}

\pagerange{\pageref{firstpage}--\pageref{lastpage}} \pubyear{2006}

\maketitle

\label{firstpage}

\begin{abstract}

We present radio observations of comet 9P/Tempel 1 associated with the Deep 
Impact spacecraft collision of 2005 July 4. Weak 18-cm OH emission was detected 
with the Parkes 64-m telescope, in data averaged over July 4 to 6, at a level
of $12 \pm 3$ mJy km/s, corresponding to OH production rate
$2.8 \times 10^{28}$ molecules/second (Despois et al. inversion model,
or $1.0 \times 10^{28}$ /s for the Schleicher \& A'Hearn model).
We did not detect the HCN 1-0 line with
the Mopra 22-m telescope over the period July 2 to 6. The $3 \sigma$ limit of
0.06 K km/s for HCN on July 4 after the impact gives the limit to the HCN 
production rate of $ < 1.8 \times 10^{25}$ /s. 
We did not detect the HCN 1-0 line, 6.7 GHz CH$_3$OH line or
3.4-mm continuum with the Australia Telescope Compact Array (ATCA) 
on July 4, giving further limits on any small-scale structure due to an 
outburst. The $3 \sigma$ limit on HCN emission of 2.5 K km/s from the ATCA 
around impact corresponds to limit $ < 4 \times 10^{29}$ HCN molecules released 
by the impact.

\end{abstract}


\section{Introduction}

The NASA mission ``Deep Impact'' encountered Comet 9P/Tempel 1 
around 05:44 UT on 2005 July 4, with the high velocity collision of the 
impactor with the comet. The impactor
had mass 370 kg and hit with a relative velocity of 10.2 km/s, so it
was expected to excavate a hole to the depth of several tens of metres, and 
around 100 m in diameter.
There was a coordinated
international campaign of ground-based and satellite observations 
(Meech et al. 2005a, 2005b) to observe the impact event, to complement the 
observations from the flyby part of the Deep Impact mission, and to 
follow changes in the comet as new activity developed after the impact.

Comets are often described as ``the most pristine material in the solar system''
but the material we observe from the gas and dust in the coma has been 
chemically and physically processed by both solar heating and radiation.
Even the surface of the nucleus has been modified from the original,
presumably amorphous (Bar-Nun \& Laufer 2003), 
ices. The dust and volatiles (similar to the icy grains in the ISM)
are processed
to a surface crust, concurrent with the loss of volatiles, crystallisation 
and the formation of a dust mantle.
One of the major aims of the Deep Impact mission was to study the 
`pristine' material below
the surface by exposing it in the crater formed by the impactor.  

The major chemical constituents of comets are H$_2$O (which dissociates to OH),
CO$_2$, CO and CH$_3$OH (methanol). Other important minor constituents, that 
can be observed at millimetre/submillimetre wavelengths are H$_2$CO 
(formaldehyde), CS, H$_2$S, HCN, HNC and CH$_3$CN. The relative abundances are 
found to vary considerably between different comets (Biver et al. 2002).  

Comet 9P/Tempel 1 is a typical Jupiter-family periodic comet, chosen largely
for its favourable orbit for the spacecraft encounter (A'Hearn et al. 2005a). 
At encounter, it was
close to perihelion, with heliocentric distance 1.51 AU and geocentric
distance 0.89 AU. 
The expected water release was $10^{28}$ 
molecules/second, before the effect of the impact (Lisse et al. 2005).

The spacecraft impact was a ``successful'' experiment releasing 
more than $10^6$ kg
of dust (Meech et al. 2005b), leading to changes monitored by the
flyby spacecraft (A'Hearn et al. 2005b) and the international campaign
(Meech et al. 2005b) over a wide range of wavelengths and techniques.  

\section[]{Observations}

The timing of the impact/flyby was  chosen to allow simultaneous (redundant)
NASA Deep Space Network (DSN) ground-station coverage from Goldstone 
(California) and Tidbinbilla 
(Australia). For the radio part of the international campaign, this meant
Australia was geographically well-placed to monitor changes in the first few
hours after impact on July 4. We made radio observations with the Parkes 64-m 
telescope to monitor OH, the Mopra 22-m telescope to monitor HCN (and CS) and 
the Australia Telescope Compact Array to detect any small-scale HCN, 3-mm
continuum or 6.7 GHz CH$_3$OH maser.

\subsection{Parkes OH observations}

Parkes observations were made of the 1667.3590 and 1665.4018 MHz OH lines
simultaneously, with frequency switching of the 8 MHz bandwidth between 
centre 1665.9 and 1666.9 MHz. The observations were made on 3 days, 
July 4, 5 and 6, with the on-source UT 
times and total integration time as given in Table \ref{pks_results}.
We used frequency switching (10 second period) rather than position switching 
for bandpass
calibration, to maximise the on-source integration time. We used a 
correlator configuration with 8192 channels, giving 0.95 kHz or 0.176 km/s 
channels. The halfpower beamwidth at 1.66 GHz was 14 arcmin.
The comet was tracked using the proper motion from the ephemeris, updated
every hour or so. The observing setup was checked by observations of the OH
maser lines in evolved stars VY CMa, V Ant, R Crt and W Hya.

The data were reduced with the ATNF SPC package \\
(http://www.atnf.csiro.au/computing/software/spc.html) which handles
the RPFITS format output of the ATNF correlators, and allowed flexible, 
albeit labour intensive, processing of the buffers for the non-standard 
frequency-switched observations. The raw spectra had four quadrants, with 
spectra centred at both 1665.9 and 1666.9 MHz and two circular polarisations 
of the H-OH receiver. The bandpass correction was made by using the 1665.9 MHz
data as signal, and the 1666.9 MHz data as reference in the signal-reference 
quotient, and the two polarisations were averaged. The data were folded
(copied, shifted by the 1 MHz frequency shift, inverted and averaged) to combine
the frequency shifted spectra. There were not regular standing waves in the 
spectra, but the baselines were still not very flat after the frequency 
switching calibration. The residual bandpass baseline was removed with 
a fifth-order polynomial.
Separate spectra were extracted for the 1667.3590 and 
1665.4018 MHz transitions, for each day of observation, corrected for 
different rest frequencies onto a geocentric velocity scale. The spectra were 
also shifted onto the velocity scale of the comet, using the geocentric
velocity of the comet given in the JPL Horizons 
(http://ssd.jpl.nasa.gov/horizons.html) ephemeris (around 9 km/s). 

\subsection{Mopra HCN and CS observations}

Mopra observations were made of the 88.63 GHz HCN $1-0$ (hyperfine triplet
88630.4157, 88631.8473 and 88633.936 MHz) and 97.98095 GHz CS $2-1$ lines, 
as given in Table \ref{mop_results} on five days, July 2 -- 6.  
We used a correlator configuration with two linear polarisations each 
32 MHz bandwidth and 1024 channels, 
giving 31 kHz or 0.106 km/s channels for HCN and 0.096 km/s for CS.
Position switching mode was used to get good bandpass baseline. 
The halfpower beamwidth was 35 arcsec, and the comet was tracked using the 
positions from the JPL Horizons ephemeris. Pointing was 
checked with SiO masers after retuning the second polarisation. 
The pointing rms was around 9 arcsec, so this dominated over any residual
errors in the ephemeris tracking. The combination
of on-off switching, and pointing checks, meant that the on-source integration
time was somewhat less than 50 percent of the observing time eg. 3.1 hours
on July 4. The weather on July 2 -- 4 was good for 3-mm observing, as shown
by the $T_{sys}$ in Table \ref{mop_results}, with the before impact July 4 data
at low elevation, but deteriorated on July 5 and 6 giving higher $T_{sys}$. 

The data were also reduced with the ATNF SPC package, with the DFM (Data From Mopra)
script \\
(http://www.phys.unsw.edu.au/astro/mopra/software.php) used to automate
the on-off position switching quotients and polarisation averaging. The spectra
were combined for each day in geocentric velocity and 
also shifted onto the velocity scale of the comet, for fitting and combining
different days.

\subsection{ATCA HCN and continuum observations}

Australia Telescope Compact Array (ATCA) observations of the HCN $1-0$ line 
were made on July 4, using the H75 array. This array provided two-dimensional
{\it u,v} - coverage for the observations near the equator, and short baselines.
The array at 3-mm uses five (of the six) 22-m telescopes, with primary beam 35 
arcsec.  The correlator configuration was 
FULL\_16\_256-64 with the first IF having 16 MHz bandwidth and 256 
channels each 
62.5 kHz or 0.21 km/s and the second IF 64 MHz bandwidth and 64 channels for 
calibration and continuum observations. The phase and pointing 
centres were tracked using an ephemeris file generated from the JPL Horizons 
ephemeris. The strong 3-mm sources 0537-441 and 1253-055 (3C279) were used
as setup sources and primary calibrators. The secondary, phase calibrator was 
1334-127. The pointing was updated using 1334-127 every hour or 
so, and the system temperature measured every half hour.

The observations were reduced with the MIRIAD package 
(Sault, Teuben \& Wright 1995; 
http://www.atnf.csiro.au/computing/software/miriad/) including
the specific routines for 3-mm ATCA observations, such as corrections for 
system temperature, gain as a function of elevation and antenna position errors.
The primary flux calibration used sources 0537-441, 1253-055 and 1830-211
and flux values from ATCA monitoring of strong, albeit, variable calibrators
relative to the flux of the planets. We did not observe planetary flux 
calibrators during our run, as the planets were either 
resolved or not in the accessible right ascension range. The synthesised beam 
size is 6.7 $\times$ 4.8 arcsec$^2$ at position angle 112 deg.

\subsection{ATCA CH$_3$OH observations}

The antenna no. 6 of ATCA, which is not equipped with the 3-mm receiver,
was used to observe the $5_1-6_0$~A$^+$ methanol (CH$_3$OH)
transition at 6668.5128 MHz simultaneously with the HCN 1-0 observations
by the rest of the array on July 4. The broad band signal was fed into the
Long Baseline Array data acquisition system (LBA DAS) configured to
digitize 4~MHz of bandwidth. This digital signal was recorded
on a computer hard disk using the capture card manufactured by the
Mets\"ahovi Radio Observatory (the VSIB card). The autocorrelation
spectra were calculated off-line for each 10 seconds interval using the
software supplied with the capture card. The mode used in software
correlation split the 4~MHz bandwidth selected
by DAS into 1024 spectral channels, providing a 0.18~km~s$^{-1}$ spectral
resolution. Two orthogonal linear polarizations were recorded separately
and averaged during the data processing. The pointing centre was updated
once per hour (at 30 minutes of each UT hour) using the JPL Horizons
ephemeris. At the given frequency the full width at half maximum of the beam
was 7.2 arcmin. Although the antenna constantly tracked
the comet, the hardware stopped recording periodically for a short period
of time (probably to flush the buffers). This resulted in 10$-$30\% loss
of integration time.

The data processing was done using our own software. It included a weighted
averaging to form 5 minute scans of data, a high-order polynomial
baselining of each polarization for each scan, followed by the weighted
averaging of individual polarizations and scans. Because the noise diode
was not used in this non-standard setup, the rms in the spectrum was used
as a measure of T$_{sys}$ to calculate the weights of individual spectra for
averaging.
These weights appeared to stay almost the same (difference less than 0.1\%)
througout the observations with the constant factor of about 1.8 between
different polarizations. The flux scale was established by observing
a test maser source G9.62-0.20E. This source exhibits periodic
flares, with the date of our observations falling into a quiescent stage
characterized by the flux density about 4700~Jy (Goedhart et al., 2003).
The accuracy of the absolute flux density scale is determined by the
``light curve'' of Goedhart et al. (2003) and is about 10\%. The radial
velocity of this maser allowed it to be observed in the same 4~MHz band with
the same setup of the conversion chain (local oscillators), as was used 
for the comet. Therefore, the maser was observed a few times to check the 
system and to test the conversion chain setup, which determines the relation 
between the spectral channels and frequencies.

\section[]{Results}

\subsection{Parkes OH results}

\begin{figure}
\includegraphics[angle=-90,width=8.5cm]{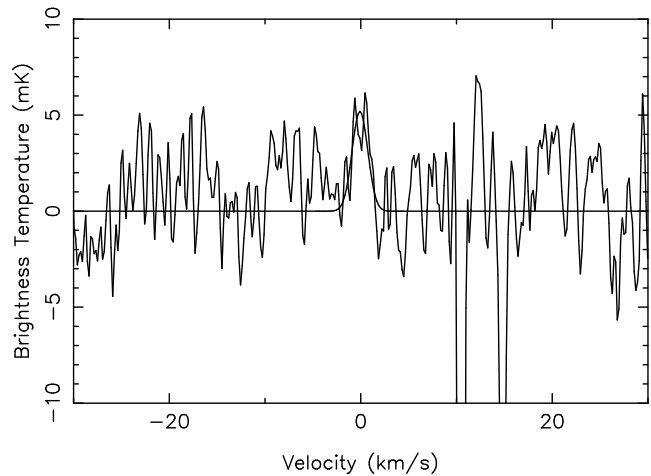}
\caption{Spectrum of the OH 1667-MHz line in comet 9P/Tempel 1, integrated over
3 days of Parkes observations, 2005 July 4 -- 6. The spectra have been 
combined after correcting for the geocentric velocity of the comet, so the 
velocity scale is relative to the comet ephemeris.}
\label{OH_1667_all}
\end{figure}

The Parkes OH spectra were averaged over each of the three days of observations,
and combined for the three days. A summary of the results is given in Table 
\ref{pks_results}. We find no significant emission at the expected velocity
(zero in our cometocentric velocity) for the 1667 or 1665 MHz OH lines on July 
4 (the day of impact) or July 5 (one day after). We do find weak peaks for 
1667 MHz line, but not the 1665 MHz line, on July 6 and the data for all 
three days combined. The spectrum of the 1667 MHz line for the three days
combined is shown in Fig. \ref{OH_1667_all}, with three point Hanning 
smoothing applied, and with the gaussian-line fit over-plotted. 

\begin{table*}
\caption{Parkes 1667 and 1665 MHz OH results}
\begin{tabular}{lccccccccccc}
\hline
Date & UT range & Int. time  & Mean          & \multicolumn{3}{c}{RMS} &
\multicolumn{3}{c}{Flux or 3 $\sigma$} & $Q_{OH}$ & $N_{OH}$ \\
     &          & (hours)    & $T_{sys}$ (K) & \multicolumn{3}{c}{$T_a$ (mK)} &
\multicolumn{3}{c}{limit (mJy km/s)} & ($10^{28}$ /s) & ($10^{34}$) \\
     &          &          &      & 1667 & 1665 & mean & 1667 & 1665 & mean 
& 1667 & 1667 \\
\hline
Jul 4 & 04:55 -- 12:50 & 7.39 & 26.4 & 4.4 & 4.9 & 3.3 & 
$<$ 12     & $<$ 14 & $<$ ~9 & $<$ 3.0 & $<$ 1.3 \\
Jul 5 & 05:09 -- 12:54 & 5.95 & 26.6 & 6.5 & 6.0 & 4.3 & 
$<$ 18     & $<$ 17 & $<$ 12 & $<$ 4.3 & $<$ 2.0 \\
Jul 6 & 04:50 -- 12:55 & 7.96 & 26.9 & 5.1 & 5.3 & 3.7 & 
16 $\pm$ 5 & $<$ 15 & 12 $\pm$ 3 & 3.9 & 1.7 \\
       &               &      &      &     &     &    &    &    &   &     \\
Jul 4 -- 6 &            & 21.3 & 26.7 & 3.1 & 3.2 & 2.3 & 
12 $\pm$ 3 & $<$ ~9 &  12 $\pm$ 3 & 2.8 & 1.3 \\
\hline
\end{tabular}
\label{pks_results}
\end{table*}

There are some 
problems with the spectra. There are interference lines, that appear due to the
frequency switching as positive peaks with two negative peaks half the 
intensity 1 MHz on either side. In Fig. \ref{OH_1667_all}, these appear as 
two strong negative features. Also, the spectral baselines are not quite flat,
with some bumps, rather than simply thermal noise. This is perhaps not 
surprising, as with such long integrations (21.3 hours for all three days 
combined) we do find the RMS noise (Table \ref{pks_results}) has decreased
with the expected $t^{-1/2}$ to a factor $10^{-4}$ of the bandpass level 
($T_{sys} \sim 27$~K). It is quite a stringent requirement for the bandpass to 
be smooth to this $10^{-4}$ level, after frequency switch calibration.

The fitted emission in the 1667 MHz OH line integrated over the three days 
is peak $5.3 \pm 1.2$ mK, velocity $0.0 \pm 0.2$ km/s and FHWM $2.0 \pm 0.6$ 
km/s. Note that the uncertainty in the fitted peak, obtained with the fit
in the SPC package, is smaller than the RMS of the individual (0.176 km/s) 
channels, since the line is resolved over many channels. We do consider the line
significant at the 3 $\sigma$ level taking this into account. 
The fitted emission line for July 6 is $6.9 \pm 2.1$ mK, velocity $-0.1 \pm 0.2$
km/s and FWHM $2.1 \pm 0.5$ km/s.

The integrated flux density of the line is given in Table \ref{pks_results}
(in mJy km/s) using the beam efficiency 0.7 and sensitivity 1.5 Jy/K from the
Parkes RadioTelescope Users Guide. We also quote 3 $\sigma$ upper limits
for the spectra without line detections, assuming limits to the fits for similar
line width (2.0 km/s) as the weak detections. We use integrated flux in mJy
rather than brightness in mK for the Parkes OH, as the scale of emission is 
smaller than the 14 arcmin beam. 

We have also averaged the spectra for the
1665 and 1667 MHz lines, with the results shown in table \ref{pks_results}.
The 1665 and 1667 MHz OH lines have statistical weights 5 and 9 so this should
slightly improve the signal to noise ratio.
This does reduce the RMS noise, and we do detect the weak line at the expected 
velocity for the July 6 data and the data averaged over the three days.
However, the gaussian fits to the lines are probably less robust that the fits 
to the 1667 MHz data, as there is a greater problem with interference lines in
the combined 1667 and 1665 MHz data.

\subsection{Mopra HCN and CS results}

The Mopra HCN spectra were averaged over each of the five days of observations,
except for the day of impact July 4, where the data were kept as before and 
after impact spectra. Since the observations were only of order 1 hour on-source 
integration time, except for July 4, we also combined data before impact for 
July 2 -- 4, and after impact July 5 -- 6. 
A summary of the results is given in Table \ref{mop_results}.
The CS was only observed at Mopra on July 3.

We did not detect either HCN or CS to quite low limits. The 3 $\sigma$ limits
in Table \ref{mop_results} were calculated from the RMS $T_a$ in the spectra, 
assuming 
nominal gaussian line half-power width 2.0 km/s. For HCN, to account 
for the three hyperfine components in the nominal 1:5:3 ratio, 
we can improve the signal to noise by
shifting and averaging the two strongest hyperfine components and 
mutiplying by (9/8). This gives 
an extra factor $(9/8)\sqrt{2} = 1.59$ in the limit (compared to factor
(9/5) = 1.8 for simply taking the limit on the strongest hyperfine component
and correcting for the extra weaker components).
The limits are in $T_{MB}$, so there is a further factor of dividing by 
0.49 for the main beam efficiency (Ladd et al. 2005).

\begin{table*}
\caption{Mopra HCN 1-0 and CS 2-1 results. The HCN data are combined over
each day of observation, except for the July 4 data which is averaged separately
before and after the impact event. The data are further combined July 2 -- 4 
before impact and July 5 -- 6 after impact, and all five days data.}
\begin{tabular}{lccccccc}
\hline
Date & UT range & Int. time  & Mean          & RMS       & 3 $\sigma$ limit &
 $Q_{HCN}$ & $N_{HCN}$ \\
     &          & (hours)    & $T_{sys}$ (K) & $T_a$ (mK) & (K km/s) &
($10^{25}$ /s) & ($10^{30}$) \\
\hline
HCN           &               &      &     &    &          &  \\
Jul 2        & 08:24 -- 10:54 & 1.17 & 209 & 17 & $<$ 0.11 & $<$ 3.3 & $<$ 0.6 \\
Jul 3        & 07:20 -- 12:52 & 1.00 & 185 & 15 & $<$ 0.10 & $<$ 2.8 & $<$ 0.5 \\
Jul 4 before & 03:58 -- 05:54 & 0.57 & 246 & 29 & $<$ 0.20 & $<$ 5.7 & $<$ 1.1 \\ 
Jul 4 after  & 06:10 -- 13:28 & 2.57 & 190 & ~9 & $<$ 0.06 & $<$ 1.8 & $<$ 0.3 \\  
Jul 5        & 04:09 -- 06:40 & 0.83 & 243 & 22 & $<$ 0.15 & $<$ 4.4 & $<$ 0.8 \\
Jul 6        & 04:26 -- 07:11 & 0.88 & 272 & 24 & $<$ 0.16 & $<$ 4.7 & $<$ 0.9 \\ 
              &               &      &     &    &          &          &     \\
Jul 2 -- 4 before &            & 2.73 & 204 & 10 & $<$ 0.07 & $<$ 1.9 & $<$ 0.4 \\
Jul 5 -- 6 after &             & 1.72 & 257 & 15 & $<$ 0.10 & $<$ 3.0 & $<$ 0.6 \\
Jul 2 -- 6 all &               & 7.02 & 206 & ~6 & $<$ 0.04 & $<$ 1.1 & $<$ 0.2 \\
              &               &      &     &    &           &        &     \\ 
CS            &               &      &     &    &          &         &     \\
Jul 3        & 08:37 - 09:51 & 0.50 & 163 & 20 & $<$ 0.08 & -        &   -  \\
\hline
\end{tabular}
\label{mop_results}
\end{table*}

\subsection{ATCA HCN and continuum results}

The ATCA HCN 1-0 and 3.4-mm continuum results are summarised in 
Table \ref{atca_results}. We did not detect either the HCN in the 
line data cubes or the continuum emission. As well as combining all the data 
over the 8 hours of observations, we have broken up the ATCA data into roughly
two-hour sections because we would expect 
compact (few arcsec) structure caused by the spacecraft impact to change over 
timescales of an hour or so. Two hours is about the shortest period that can 
be imaged with adequate $\it u,v$-coverage with the H75 array. The last two hour 
section has high noise, as the comet was getting low in the sky and data were 
lost from antenna shadowing of the shortest EW baselines.

The RMS in the continuum images and line data cubes are given in mJy/beam
for the 6.7 $\times$ 4.8 arcsec$^2$ synthesised beam. The point source continuum
flux limits can be taken as 3$\times$ RMS, for the nucleus. 
The dust cloud released by the impact should be a point source for the ATCA 
beam for several hours, given the dust velocity of 200 m/s reported by 
Meech et al. (2005b), and the velocity of larger particles seen in the 
millimetre wavelength is probably even less.
The limit for 
the extended coma emission from the normal steady comet outflow 
is somewhat larger, with the expected 
$r^{-2}$ dependence of dust density giving increasing flux when
summed over larger and larger scales (eg. de Pater et al. 1998).

The spectral line limits were calculated, similar to the Mopra limits, assuming
fitting errors for 2.0 km/s FWHM lines of the HCN triplet, and converted
from mJy/beam to brightness temperature $T_B$ with 
$T_B = (S/\Omega)(\lambda^{2}/2 k)$ where $\Omega$ is beam solid angle.

\begin{table*}
\caption{ATCA HCN 1-0 line and 3.4-mm continuum results.}
\begin{tabular}{lcccccc}
\hline
Date & UT range &  RMS  & 3 $\sigma$ limit & RMS &  $Q_{HCN}$ & $N_{HCN}$ \\
     &          & line  & line             & continuum  &    & \\
     &          & (mJy) & (K km/s)  & (mJy) & ($10^{25}$ /s) & ($10^{30}$) \\
\hline     
Jul 4  & 05:30 -- 13:20 & ~48  & $<$ 1.2 & 1.6 & $<$ ~6 & $<$ 0.2 \\
        &               &      &         &     &     &     \\ 
Jul 4  & 05:30 -- 07:30 & ~99  & $<$ 2.5 & 3.1 & $<$ 12 & $<$ 0.4 \\
Jul 4  & 07:30 -- 09:30 & ~75  & $<$ 1.9 & 2.2 & $<$ ~9 & $<$ 0.3 \\ 
Jul 4  & 09:30 -- 11:20 & ~77  & $<$ 1.9 & 2.5 & $<$ ~9 & $<$ 0.3 \\  
Jul 4  & 11:20 -- 13:20 & 155  & $<$ 3.9 & 5.5 & $<$ 18 & $<$ 0.6 \\
\hline
\end{tabular}
\label{atca_results}
\end{table*}

\subsection{ATCA CH$_3$OH results}

The ATCA CH$_3$OH $5_1-6_0$~A$^+$ results are summarized in Table 
\ref{atca_results_meth}.
Similarly to the HCN 1-0 and 3.4-mm continuum results, in addition to
combining all the data, four shorter (roughly two-hour) intervals have also
been examined. We did not detect any methanol emission in either of
these time intervals. The limits were calculated the same way as for Mopra
data, assuming that 1~K of the main beam brightness temperature of a single
ATCA antenna is equivalent to 6.8~Jy.

\begin{table}
\caption{ATCA CH$_3$OH 5$_1-6_0$~A$^+$ results}
\begin{tabular}{lcccc}
\hline
Date   & UT range         & Int. time &  RMS  & 3 $\sigma$ limit \\
       &                  & (hours)   & (mJy) &  (mK km~s$^{-1}$) \\
\hline
Jul 4 & 05:45~$-$~13:14  & 5.94      &  180  & $<$ 4.2  \\ 
\\
Jul 4 & 05:45~$-$~07:30  & 1.32      &  360  & $<$ 8.9  \\ 
Jul 4 & 07:30~$-$~09:30  & 1.60      &  340  & $<$ 9.4  \\ 
Jul 4 & 09:30~$-$~11:14  & 1.40      &  350  & $<$ 8.0  \\ 
Jul 4 & 11:14~$-$~13:14  & 1.62      &  350  & $<$ 8.5  \\ 
\hline
\end{tabular}
\label{atca_results_meth}
\end{table}

\section{Discussion} 

There was an extensive campaign of observations of comet 9P/Tempel 1 
(Meech et al. 2005a, 2005b) associated with the Deep Impact mission, from the
ground and spacecraft, over a wide range of wavebands. The results
presented here will be most useful when combined with other complementary data
into physical models.
These data include (but are not restricted to) other 18-cm OH 
observations, from
the Green Bank Telescope (GBT) and Nan\c{c}ay, and millimetre HCN, CS and 
CH$_3$OH 
observations from the IRAM-30m, CSO-10m and JCMT, including higher transitions.
In the steady state, the models typically assume constant velocity outflow
from the nucleus (which would give radial dependence of density 
$n(r) \propto r^{-2}$ and column density integrated along the line of sight
$N(r) \propto r^{-1}$), modified by the time-dependent formation and 
destruction of the species of interest. 

For lines such as HCN 1-0 the 
conversion between integrated line intensity and column density (and production 
rate) is relatively straight-forward, using standard radiative transfer 
equations (e.g. Crovisier et al. 1987), but for the 18-cm OH lines there is 
additional complexity (e.g. Crovisier et al. 2002) with 
UV-pumping dependent on the heliocentric velocity, leading to inversion
or anti-inversion of the levels (Despois et al. 1981; Schleicher \& A'Hearn
1988). Taking into account the variety of maser transitions and pumping
regimes observed in the interstellar medium, the case of the methanol
emission is likely to be even more complex.

\subsection{Production rates and number of molecules}

We present here rough estimates of molecule production rates, or number of 
molecules, from our observations.

For OH, we have used the conversion between observed line flux (in mJy km/s)
and derived production rate ($Q_{OH}$ in molecules/s) from the work of
Crovisier et al. (2002) with Nan\c{c}ay, scaling for inversion factor, 
heliocentric distance, background brightness temperature and beam area.
We quote $Q_{OH}$ in Table \ref{pks_results} and also calculate
the number of OH molecules ($N_{OH}$) within the Parkes beam (7 arcmin
or 272 000 km radius). We used the inversion factor 0.03 appropriate
for the Despois et al. (1981) model, but note that the model of 
Schleicher \& A'Hearn (1998) has inversion factor 0.08, which
would decrease the estimates of $Q_{OH}$ and $N_{OH}$ by the factor 3/8.
Given the discrepancies between the models at this small inversion factors
(near a zero crossing) around heliocentric velocities near zero, the 
estimates of $Q_{OH}$ and $N_{OH}$ are subject to quite a large systematic 
error.

From the Mopra HCN flux limits, we derived the limits on HCN production rate
($Q_{HCN}$) in Table \ref{mop_results}, by scaling from the predicted flux 
for the IRAM 30-m from Meech et al. (2005a), for the different beamsizes,
and corrected for the ratios of the different HCN levels as observed by Biver 
et al. (2005a). The limits on the number of HCN molecules within the Mopra 
35 arcsec beam (radius 11 300 km) was calculated from the production rate,
on the simple constant velocity outflow model.

We similarly derived limits on HCN production rates in Table \ref{atca_results}
from the ATCA HCN flux limits, and limits on the number of HCN molecules within 
the ATCA beam (radius 1800 km). 

Note that expressed in brightness temperature, the Mopra limits on HCN are 
much lower than the ATCA limits, as 
Mopra has a much larger beam solid angle $\Omega$, and
this corresponds to somewhat lower limits with Mopra on the HCN production rate.
However, the ATCA results
are sensitive to the small-scale structure that might have been expected 
from the spacecraft impact causing an outburst. 
In limiting such an outburst, the most appropriate 3 $\sigma$ limit is
probably $N_{HCN} < 4 \times 10^{29}$ molecules 
from the ATCA over the period 05:30 - 07:30 UT,
since the compact structure would expand out of the ATCA beam 
over the period of an hour or so. 
Similar limits on $N_{HCN}$ are obtained for the Mopra and ATCA data 
averaged over the time after impact on the day Jul 4, although when averaging
over these longer times material released by the impact would leave the beam 
area.

For the radio continuum, we used the simple model of Jewitt \& Mathews (1997)
and the 3 $\sigma$ limit of 4.8 mJy for Jul 4. 
We assumed temperature 300 K, 
as the rough average of the measured temperature range 260 to 329 K
(A'Hearn et al., 2005b), giving black-body cross-sectional area
less than $1.2 \times 10^{9}$ m$^{2}$, as the limit on the coma dust.
The thermal emission from the nucleus would be much less, given the mean radius
3.0 km (A'Hearn et al., 2005b). 
To convert the cross-section to mass, we assumed from Jewitt \& Matthews (1997) 
dust
opacity $\kappa(\lambda) = \kappa(\lambda_{o}) (\lambda/\lambda_{o})^{-\beta}$
with $\kappa$(1 mm) = 0.05 m$^{2}$ kg$^{-1}$, and 
$\beta = 0.9$.
This gave dust mass less than $7 \times 10^{10}$ kg within the ATCA beam
(radius around  1800 km). Assuming a dust velocity 200 m/s, from 
Meech et al. (2005b), the timescale for dust moving out of the beam is several
hours, so this mass can be taken as a limit on the mass released by
the spacecraft impact, or alternatively converted to a limit on the steady
outflow of $8 \times 10^{6}$ kg/s. These limits are not very restrictive.
We also note that the opacity $\kappa$ is not well known; 
for example, Altenhoff et al. (1999) argue for a value 
$\kappa$(1 mm) = 7.5 m$^{2}$ kg$^{-1}$ which would decrease the dust mass
limits by a factor 150.

\subsection{Comparison with other results}

Pre-impact 18-cm OH observations were made at Nan\c{c}ay by Crovisier et al. 
(2005a)
and Arecibo by Howell et al. (2005) from 2005 March to June, when the line was
in strong absorption with large negative inversion parameter $i$. The average
spectrum over Mar. 20 to Apr. 14 from Nan\c{c}ay observations (Crovisier 
et al. 2005a)
had area $12 \pm 2$ mJy km/s, corresponding to OH production rate 
$0.4 \times 10^{28}$ molecules/second. The OH production rate increased, 
as would
be expected, as the comet approached perihelion. Schleicher \& Barnes (2005)
estimated $0.6 \times 10^{28}$ molecules/second for June 9 from optical 
narrow-band photometry. Post-impact OH observations were also made
at Nan\c{c}ay by Biver et al. (2005b) and the GBT
by Howell et al. (2005). The GBT observations over the period July 4 to 11 
showed
variability with a mean integrated area of 18 mJy km/s for this period
(Howell et al. 2005). This is in good agreement with our Parkes OH results
(Table \ref{pks_results}) of $12 \pm 3$ mJy km/s for the overlapping period
July 4 -- 6, given; \\
a) the variability of water emission, as determined by
the SWAS observations (see below) and \\
b) the likelihood of extra OH line variability due to
the conditions of excitation. \\
Our estimate of OH production rate over July 4 -- 6 of $2.8 \times 10^{28}$ /s
with Despois et al. (1981) inversion factor, or $1.0 \times 10^{28}$ /s
with Schleicher \& A'Hearn (1998) inversion factor, indicates some
increase in the OH production rate in the few days after impact.

The water production rate from 9P/Tempel 1 around the date of impact
was monitored in the 557 GHz line by the SWAS (Bensch et al. 2005a, 2005b)
and Odin (Biver et al. 2005b) satellites. The SWAS observations (Bensch et al.
2005b) show water production between 2005 June 5 and July 9 varying
by a factor of three from $0.4 \times 10^{28}$ to $1.3 \times 10^{28}$ 
molecules/second, but no statistically significant increase following the 
impact. They found water production rate for the three days following impact
was $(0.66 \pm 0.15) \times 10^{28}$ molecules/second, with a $3 \sigma$ limit
of $9 \times 10^{32}$ molecules released in an outburst, and less than a factor
of two increase in water production due to a new active area.
Before the impact event, it was suggested that the crater could lead to
a new focus for out-gassing which would persist (A'Hearn et al. 2005a), 
but this was not observed (Meech et al. 2005b).

Near infrared spectroscopy with the Keck-2 telescope by Mumma et al. (2005)
indicated an increase in water production rate from $1.2 \times 10^{28}$
molecules/second on June 3 and $1.0 \times 10^{28}$ molecules/second on July 4
before impact to $(1.7, 1.7, 2.1, 2.0) \times 10^{28}$ molecules/second
measured after impact on three short periods on July 4 and one on July 5.

Our results for production rates of OH (the daughter product of the water) 
are broadly consistent with these results, but there are probably some 
systematic differences in calculated production rates between the different
techniques due to model assumptions.

From observations with the ESA Rosetta spacecraft, K\"{u}ppers et al. (2005) 
derive
a total water vapour output of $1.5 \times 10^{32}$ molecules due to the impact.
This is much less than our upper limit of $ < 1.3 \times 10^{34}$ OH
molecules within the Parkes beam on July 4.
K\"{u}ppers et al. derive a large dust/ice ratio of greater than one by mass,
corresponding to a dust mass $ > 5 \times 10^{6}$ kg released by the impact.
Our limit on dust mass from the 3.4-mm radio continuum of 
$ < 7 \times 10^{10}$ kg in the ATCA beam is, as noted earlier, not very 
restrictive.

Pre-impact millimetre-wave observations of HCN and other molecules
were made with the IRAM-30m telescope by Biver et al. (2005a) giving integrated
intensity of the HCN $1-0$ line of $0.031 \pm 0.004$ K km/s over May 4.8 to 9.0. 
These HCN observations showed
periodic variability in HCN production rate over the range 
$(0.5 - 1.0) \times 10^{25}$ molecules/second with period 1.7 days associated 
with the nucleus rotation. They did not detect CS. Our 
$3 \sigma$ limits to the HCN 1-0
from the Mopra observations (Table \ref{mop_results}) combined over the 
July 2 - 6 
period ($ < 1.1 \times 10^{25}$ /s) are close to this level and our results 
for the few
days post-impact ($ < 1.8 \times 10^{25}$ /s on July 4 and 
$ < 3.0 \times 10^{25}$ /s on July 5 -- 6)
rule out any post-impact increase greater than about a factor of two.
Mumma et al. (2005) report HCN production rates $ 2.1 \times 10^{25}$ /s 
before impact (June 3) and $ 3.6 \times 10^{25}$ /s 
after impact (July 4), with constant HCN to H$_2$O ratio
of $0.2$ \%. This indicates that the increase due to the impact was small.
However, these rates from Mumma et al. are a factor of two higher than
our limit for July 4, and the pre-impact level from Biver et al. (2005a),
which may indicate systematic errors due to model assumptions in translating
HCN line fluxes to production rates. 

Our limit on the total HCN released
by the impact (ATCA data over 05:30 to 07:30 UT) of $ < 4 \times 10^{29}$ 
molecules is close to the value of $ 3 \times 10^{29}$ molecules
(0.2 \% of water amount $ 1.5 \times 10^{32}$) from Mumma et al.

Methanol (CH$_3$OH) is quite abundant in
comets (e.g., Crovisier et al., 2005b). Biver et al. (2005a) found a
production rate of methanol in the comet 9P/Tempel~1 of
about (1$-$2)$\times10^{26}$ molecules/second. However, most methanol
transitions observed so far are seen in thermal emission in the interstellar 
medium.
The $5_1-6_0$~A$^+$ transition observed in our study is the brightest known
maser transition. It has a relatively low spontaneous decay rate
(about 1.5$\times10^{-9}$~s$^{-1}$ according to the Hamiltonian of
Mekhtiev, Godfrey \& Hougen (1999) and usually has to be inverted 
(i.e. be a maser)
to produce a detectable emission. The physical conditions required for
this inversion to happen are studied only for the range of parameters
typical for star-forming regions, where such masers are found. 
It is worth noting, however, that Slysh (2004) has
suggested a model for interstellar masers in this transition, which
form in large comet-like objects orbiting protostars. Although, no
quantitative simulations of such model has been reported,
the non-detection reported in this paper rules it out for the conditions
typical for the Solar System. 

\section*{Acknowledgments}

We would like to thank several people for help with these non-standard 
observations:
Mike Kesteven for modifications of the Mopra control 
system to allow proper motion tracking of the comet; 
John Reynolds and Brett Preisig for help with Parkes schedules, receiver
and frequency-switching hardware;
Bob Sault for help with the ATCA schedule files; David Brodrick and Mark 
Wieringa for help with the ATCA control software;
and Steven Tingay and Craig West for help with the broad-band disk recorder 
setup.
We would like also to thank Nicolas Biver for helpful discussion, and
referee's comments.
The Australia Telescope is funded by the Commonwealth
of Australia for operation as a National Facility managed by CSIRO.



\label{lastpage}

\end{document}